

\input harvmac

\noblackbox
\baselineskip 20pt plus 2pt minus 2pt

\overfullrule=0pt



\def\fpi { f_{,i} }
\def\Fpi { F_{,i} }

\def\Fpu { F_{,u} }
\def\Fpv { F_{,v} }
\def\Apu { A_{,u} }
\def\Apv { A_{,v} }
\def\gpu { g_{,u} }
\def\gpv { g_{,v} }

\def\SW {Schwarzschild}
\def\RN {Reissner-Nordstr\"om}
\def\CS {Christoffel}
\def\MI {Minkowski}

\def\bs{\bigskip}

\def\hb{\hfill\break}
\def\qq{\qquad}
\def\bl{\bigl}
\def\br{\bigr}

\def\IR{\relax{\rm I\kern-.18em R}}

\def\np {  Nucl. Phys. }


\def\r{\rho}
\def\a{\alpha}

\def\b{\beta}

\def\G{\Gamma}
\def\d{\delta}

\def\e{\epsilon}

\def\th{\theta}

\def\m{\mu}
\def\n{\nu}

\def\l{\lambda}

\def\s{\sigma}

\def\IR{\relax{\rm I\kern-.18em R}}

\def \bd {\bar \del}

\def \z { {\bar z} }

\def \ha {{1\over 2}}

\def \ov {\over}

\def\const{{\rm const.}}



\lref\BSthree{I. Bars and K. Sfetsos, Mod. Phys. Lett. {\bf A7} (1992) 1091.}

\lref\BShet{I. Bars and K. Sfetsos, Phys. Lett. {\bf 277B} (1992) 269.}

\lref\BSglo{I. Bars and K. Sfetsos, Phys. Rev. {\bf D46} (1992) 4495.}

\lref\BSexa{I. Bars and K. Sfetsos, Phys. Rev. {\bf D46} (1992) 4510.}

\lref\SFET{K. Sfetsos, Nucl. Phys. {\bf B389} (1993) 424.}

\lref\BSslsu{I. Bars and K. Sfetsos, Phys. Lett. {\bf 301B} (1993) 183.}

\lref\BSeaction{I. Bars and K. Sfetsos, Phys. Rev. {\bf D48} (1993) 844.}

\lref\BN{ I. Bars and D. Nemeschansky, Nucl. Phys. {\bf B348} (1991) 89.}

\lref\BCR{K. Bardakci, M. Crescimanno and E. Rabinovici,
Nucl. Phys. {\bf B344} (1990) 344.}

\lref\WIT{E. Witten, Phys. Rev. {\bf D44} (1991) 314.}

 \lref\IBhet{ I. Bars, Nucl. Phys. {\bf B334} (1990) 125. }

 \lref\IBCS{ I. Bars, ``String Propagation on Black Holes'', USC-91-HEP-B3.\hb
{\it Curved Space-time Strings and Black Holes},
in Proc.
 {\it XX$^{th}$ Int. Conf. on Diff. Geometrical Methods in Physics}, eds. S.
 Catto and A. Rocha, Vol. 2, p. 695, (World Scientific, 1992).}

 \lref\CRE{M. Crescimanno, Mod. Phys. Lett. {\bf A7} (1992) 489.}

\lref\MSW{G. Mandal, A. Sengupta and S. Wadia,
Mod. Phys. Lett. {\bf A6} (1991) 1685.}

 \lref\HOHO{J.B. Horne and G.T. Horowitz, Nucl. Phys. {\bf B368} (1992) 444.}

 \lref\FRA{E. S. Fradkin and V. Ya. Linetsky, Phys. Lett. {\bf 277B}
          (1992) 73.}

 \lref\ISH{N. Ishibashi, M. Li, and A. R. Steif,
         Phys. Rev. Lett. {\bf 67} (1991) 3336.}

 \lref\HOR{P. Horava, Phys. Lett. {\bf 278B} (1992) 101.}

 \lref\RAI{E. Raiten, ``Perturbations of a Stringy Black Hole'',
         Fermilab-Pub 91-338-T.}

 \lref\GER{D. Gershon, Phys. Rev. {\bf D49} (1994) 999.}

\lref\GERexa{D. Gershon, ``Semiclassical vs. Exact Solutions of charged Black
Hole in four dimensions and exact $O(D,D)$ duality'', TAUP-2121-93,
hepth/9311122.}

\lref\GERexadual{D. Gershon, ``Exact $O(D,D)$ transformations in WZW models,
TAUP-2129-93, hepth/9312154.}

 \lref \GIN {P. Ginsparg and F. Quevedo,  Nucl. Phys. {\bf B385} (1992) 527. }

 \lref\HOHOS{ J.H. Horne, G.T. Horowitz and A. R. Steif, Phys. Rev. Lett.
 {\bf 68} (1991) 568.}

 \lref\groups{
 M. Crescimanno. Mod. Phys. Lett. {\bf A7} (1992) 489. \hb
 J. B. Horne and G.T. Horowitz, Nucl. Phys. {\bf B368} (1992) 444. \hb
 E. S. Fradkin and V. Ya. Linetsky, Phys. Lett. {\bf 277B} (1992) 73. \hb
 P. Horava, Phys. Lett. {\bf 278B} (1992) 101.\hb
 E. Raiten, ``Perturbations of a Stringy Black Hole'',
         Fermilab-Pub 91-338-T.\hb
 D. Gershon, ``Exact Solutions of Four-Dimensional Black Holes in
         String Theory'', TAUP-1937-91.}

\lref\NAWIT{C. Nappi and E. Witten, Phys. Lett. {\bf 293B} (1992) 309.}

\lref\FRATSE{E. S. Fradkin and A.A. Tseytlin,
Phys. Lett. {\bf 158B} (1985) 316.}

\lref\DASA{ S. Das and B. Sathiapalan, Phys. Rev. Lett. {\bf 56} (1986) 2664.}

\lref\CALLAN{ C.G. Callan, D. Friedan, E.J. Martinec and M. Perry,
Nucl. Phys. {\bf B262} (1985) 593.}

\lref\DB{L. Dixon, J. Lykken and M. Peskin, Nucl. Phys.
{\bf B325} (1989) 325.}

\lref\IB{I. Bars, Nucl. Phys. {\bf B334} (1990) 125.}

\lref\BUSCHER{T. Buscher, Phys. Lett. {\bf 194B} (1087) 59;
Phys. Lett. {\bf 201B} (1988) 466.}

\lref\RV{M. Rocek and E. Verlinde, Nucl. Phys. {\bf B373} (1992) 630.}
\lref\GR{A. Giveon and M. Rocek, Nucl. Phys. {\bf B380} (1992) 128. }

\lref\nadual{X. C. de la Ossa and F. Quevedo,
Nucl. Phys. {\bf B403} (1993) 377.}

\lref\duearl{B.E. Fridling and A. Jevicki,
Phys. lett. {\bf B134} (1984) 70.\hb
E.S. Fradkin and A.A. Tseytlin, Ann. Phys. {\bf 162} (1985) 31.}

\lref\ABEN{I. Antoniadis, C. Bachas, J. Ellis and D.V. Nanopoulos,
Phys. Lett. {\bf B211} (1988) 393.}

\lref\Kalofour{N. Kaloper, Phys. Rev. {\bf D48} (1993) 4658.}

\lref\GPS{S.B. Giddings, J. Polchinski and A. Strominger,
Phys. Rev. {\bf D48} (1993) 5784.}

\lref\SENrev{A. Sen, ``Black Holes and Solitons in String Theory'',
TIFR-TH-92-57.}

\lref\TSEd{A.A. Tseytlin, Mod. Phys. Lett. {\bf A6} (1991) 1721.}

\lref\TSESC{A. S. Schwarz and A.A. Tseytlin, ``Dilaton shift under duality
and torsion of elliptic complex'', IMPERIAL/TP/92-93/01. }

\lref\Dualone{K. Meissner and G. Veneziano,
Phys. Lett. {\bf B267} (1991) 33;
Mod. Phys. Lett. {\bf A6} (1991) 3397. \hb
M. Gasperini and G. Veneziano, Phys. Lett. {\bf 277B} (1992) 256. \hb
M. Gasperini, J. Maharana and G. Veneziano, Phys. Lett. {\bf 296B} (1992) 51.}

\lref\rovr{K. Kikkawa and M. Yamasaki, Phys. Lett. {\bf B149} (1984) 357.\hb
N. Sakai and I. Senda, Prog. theor. Phys. {\bf 75} (1986) 692.}

\lref\narain{K.S. Narain, Phys. Lett. {\bf B169} (1986) 369.\hb
K.S. Narain, M.H. Sarmadi and C. Vafa, Nucl. Phys. {\bf B288} (1987) 551.}

\lref\GV{P. Ginsparg and C. Vafa, Nucl. Phys. {\bf B289} (1987) 414.}
\lref\nssw{V. Nair, A. Shapere, A. Strominger and F. Wilczek,
Nucl. Phys. {\bf B287} (1987) 402.}

\lref\vafa{C. Vafa, ``Strings and Singularities'', HUTP-93/A028.}

\lref\Dualtwo{A. Sen,
Phys. Lett. {\bf B271} (1991) 295;\ ibid. {\bf B274} (1992) 34;
Phys. Rev. Lett. {\bf 69} (1992) 1006. \hb
S. Hassan and A. Sen, Nucl. Phys. {\bf B375} (1992) 103. \hb
J. Maharana and J. H. Schwarz, Nucl. Phys. {\bf B390} (1993) 3.\hb
A. Kumar, Phys. Lett. {\bf B293} (1992) 49.}

\lref\dualmargi{S.F. Hassan and A. Sen, Nucl. Phys. {\bf B405} (1993) 143.\hb
M. Henningson and C. Nappi, Phys. Rev. {\bf D48} (1993) 861.}

\lref\bv{R. Brandenberger and C. Vafa, Nucl. Phys. {\bf B316} (1989) 301.}
\lref\gsvy{B.R. Greene, A. Shapere, C. Vafa and S.T. Yau,
Nucl. Phys. {\bf B337} (1990) 1.}
\lref\tv{A.A. Tseytlin and C. Vafa,  Nucl. Phys. {\bf B372} (1992) 443.}

\lref\grvm{A. Shapere and F. Wilczek, Nucl. Phys. {\bf B320} (1989) 609.\hb
A. Giveon, E. Rabinovici and G. Veneziano,
Nucl. Phys. {\bf B322} (1989) 167.\hb
A. Giveon, N. Malkin and E. Rabinovici, Phys. Lett. {\bf B238} (1990) 57.}

\lref\GRV{M. Gasperini, R. Ricci and G. Veneziano,
Phys. Lett. {\bf B319} (1993) 438.}

\lref\KIRd{E. Kiritsis, Nucl. Phys. {\bf B405} (1993) 109.}

\lref\gpr{A. Giveon, M. Porrati and E. Rabinovici, ``Target Space Duality
in String Theory'', RI-1-94, hepth/9401139.}

\lref\slt{A. Giveon, Mod. Phys. Lett. {\bf A6} (1991) 2843.}

\lref\GIPA{A. Giveon and A. Pasquinucci, ``On cosmological string backgrounds
with toroidal isometries'', IASSNS-HEP-92/55, August 1992.}

\lref\KASU{Y. Kazama and H. Suzuki, Nucl. Phys. {\bf B234} (1989) 232. \hb
Y. Kazama and H. Suzuki Phys. Lett. {\bf 216B} (1989) 112.}

\lref\WITanom{E. Witten, Comm. Math. Phys. {\bf 144} (1992) 189.}

\lref\WITnm{E. Witten, Nucl. Phys. {\bf B371} (1992) 191.}

\lref\IBhetero{I. Bars, Phys. Lett. {\bf 293B} (1992) 315.}

\lref\IBerice{I. Bars, {\it Superstrings on Curved Space-times}, Lecture
delivered at the Int. workshop on {\it String Quantum Gravity and Physics
at the Planck Scale}, Erice, Italy, June 1992.}

\lref\DVV{R. Dijkgraaf, E. Verlinde and H. Verlinde, Nucl. Phys. {\bf B371}
(1992) 269.}

\lref\TSEY{A.A. Tseytlin, Phys. Lett. {\bf 268B} (1991) 175.}

\lref\JJP{I. Jack, D. R. T. Jones and J. Panvel,
          Nucl. Phys. {\bf B393} (1993) 95.}

\lref\BST { I. Bars, K. Sfetsos and A.A. Tseytlin, unpublished. }

\lref\TSEYT{ A.A. Tseytlin, Nucl. Phys. {\bf B399} (1993) 601.}

\lref\TSEYTt{A.A. Tseytlin, Nucl. Phys. {\bf B411} (1993) 509.}

 \lref\SHIF { M. A. Shifman, Nucl. Phys. {\bf B352} (1991) 87.}
\lref\SHIFM { H. Leutwyler and M. A. Shifman, Int. J. Mod. Phys. {\bf
A7} (1992) 795. }

\lref\POLWIG { A. M. Polyakov and P. B. Wiegman, Phys.
Lett. {\bf 141B} (1984) 223.  }

\lref\BCR{K. Bardakci, M. Crescimanno
and E. Rabinovici, Nucl. Phys. {\bf B344} (1990) 344. }

\lref\Wwzw{E. Witten, Commun. Math. Phys. {\bf 92} (1984) 455.}

\lref\GKO{P. Goddard, A. Kent and D. Olive, Phys. Lett. {\bf 152B} (1985) 88.}

\lref\Toda{A. N. Leznov and M. V. Saveliev, Lett. Math. Phys. {\bf 3} (1979)
489. \hb A. N. Leznov and M. V. Saveliev, Comm. Math. Phys. {\bf 74}
(1980) 111.}

\lref\GToda{J. Balog, L. Feh\'er, L. O'Raifeartaigh, P. Forg\'acs and A. Wipf,
Ann. Phys. (New York) {\bf 203} (1990) 76; Phys. Lett. {\bf 244B}
(1990) 435.}

\lref\GWZW{ E. Witten, \np {\bf B223} (1983) 422. \hb
K. Bardakci, E. Rabinovici and B. S\"aring, Nucl. Phys. {\bf B299}
(1988) 157. \hb K. Gawedzki and A. Kupiainen, Phys. Lett. {\bf 215B}
(1988) 119.; Nucl. Phys. {\bf B320} (1989) 625. }

\lref\SCH{ D. Karabali, Q-Han Park, H. J. Schnitzer and Z. Yang,
                   Phys. Lett. {\bf B216} (1989) 307. \hb D. Karabali
and H. J. Schnitzer, Nucl. Phys. {\bf B329} (1990) 649. }

 \lref\KIR{E. Kiritsis, Mod. Phys. Lett. {\bf A6} (1991) 2871. }

\lref\BIR{N. D. Birrell and P. C. W. Davies,
{\it Quantum Fields in Curved Space}, Cambridge University Press.}

\lref\WYB{B. G. Wybourn, {\it Classical Groups for Physicists }
(John Wiley \& sons, 1974).}

\lref\Brinkman{H.W. Brinkmann, Math. Ann. {\bf 94} (1925) 119.}

\lref\SANTA{R. Guven, Phys. Lett. {\bf 191B} (1987) 275.\hb
D. Amati and C. Klimcik, Phys. Lett. {\bf 219B} (1989) 443.\hb
G.T. Horowitz and A. R. Steif, Phys. Rev. Lett. {\bf 64} (1990) 260;
Phys. Rev. {\bf D42} (1990) 1950.\hb
R.E. Rudd, Nucl. Phys. {\bf B352} (1991) 489.\hb
C. Duval, G.W. Gibbons and P.A. Horvathy, Phys. Rev. {\bf D43} (1991) 3907.\hb
C. Duval, Z. Horvath and P.A. Horvathy, Phys. Lett. {\bf B313} (1993) 10.\hb
E.A. Bergshoeff, R. Kallosh and T. Ortin, Phys. Rev. {\bf D47} (1993) 5444.}
\lref\SANT{J. H. Horne, G.T. Horowitz and A. R. Steif,
Phys. Rev. Lett. {\bf 68} (1991) 568.}

\lref\tsecov{A.A. Tseytlin, Nucl. Phys. {\bf B390} (1993) 153;
Phys. Rev. {\bf D47} (1993) 3421.}

\lref\garriga{J. Garriga and E. Vardaguer, Phys. Rev. {\bf D43} (1991) 391.}

\lref\PRE{J. Prescill, P. Schwarz, A. Shapere, S. Trivedi and F. Wilczek,
Mod. Phys Lett. {\bf A6} (1991) 2353.\hb
C. Holzhey and F. Wilczek, Nucl. Phys. {\bf B380} (1992) 447.}

\lref\HAWK{J. B. Hartle and S. W. Hawking Phys. Rev. {\bf D13} (1976) 2188.\hb
S. W. Hawking, Phys. Rev. {\bf D18} (1978) 1747.}

\lref\HAWKI{S. W. Hawking, Comm. Math. Phys. {\bf 43} (1975) 199.}

\lref\HAWKII{S. W. Hawking, Phys. Rev. {\bf D14} (1976) 2460.}

\lref\euclidean{S. Elitzur, A. Forge and E. Rabinovici,
Nucl. Phys. {\bf B359} (1991) 581. }

\lref\ITZ{C. Itzykson and J. Zuber, {\it Quantum Field Theory},
McGraw Hill (1980). }

\lref\kacrev{P. Goddard and D. Olive, Journal of Mod. Phys. {\bf A} Vol. 1,
No. 2 (1986) 303.}

\lref\BBS{F.A. Bais, P. Bouwknegt, K.S. Schoutens and M. Surridge,
Nucl. Phys. {\bf B304} (1988) 348.}

\lref\nonl{A. Polyakov, {\it Fields, Strings and Critical Phenomena}, Proc. of
Les Houses 1988, eds. E. Brezin and J. Zinn-Justin North-Holland, 1990.\hb
Al. B. Zamolodchikov, preprint ITEP 87-89. \hb
K. Schoutens, A. Sevrin and P. van Nieuwenhuizen, Proc. of the Stony Brook
Conference {\it Strings and Symmetries 1991}, World Scientific,
Singapore, 1992. \hb
J. de Boer and J. Goeree, ``The Effective Action of $W_3$ Gravity to all
\hb orders'', THU-92/33.}

\lref\HOrev{G.T. Horowitz, {\it The Dark Side of String Theory:
Black Holes and Black Strings}, Proc. of the 1992 Trieste Spring School on
String Theory and Quantum Gravity.}

\lref\HSrev{J. Harvey and A. Strominger, {\it Quantum Aspects of Black
Holes}, Proc. of the 1992 Trieste Spring School on
String Theory and Quantum Gravity.}

\lref\GM{G. Gibbons, Nucl. Phys. {\bf B207} (1982) 337.\hb
G. Gibbons and K. Maeda, Nucl. Phys. {\bf B298} (1988) 741.}

\lref\GID{S. B. Giddings, Phys. Rev. {\bf D46} (1992) 1347.}

\lref\PRErev{J. Preskill, {\it Do Black Holes Destroy Information?},
Proc. of the International Symposium on Black Holes, Membranes, Wormholes,
and Superstrings, The Woodlands, Texas, 16-18 January, 1992.}

\lref\tye{S-W. Chung and S. H. H. Tye, Phys. Rev. {\bf D47} (1993) 4546.}

\lref\eguchi{T. Eguchi, Mod. Phys. Lett. {\bf A7} (1992) 85.}

\lref\blau{M. Blau and G. Thompson, Nucl. Phys. {\bf B408} (1993) 345.}

\lref\HSBW{P. S. Howe and G. Sierra, Phys. Lett. {\bf 144B} (1984) 451.\hb
J. Bagger and E. Witten, Nucl. Phys. {\bf B222} (1983) 1.}

\lref\GSW{M. B. Green, J. H. Schwarz and E. Witten, {\it Superstring Theory},
Cambridge Univ. Press, Vols. 1 and 2, London and New York (1987).}

\lref\KAKU{M. Kaku, {\it Introduction to Superstrings}, Springer-Verlag, Berlin
and New York (1991).}

\lref\LSW{W. Lerche, A. N. Schellekens and N. P. Warner, {\it Lattices and
Strings }, Physics Reports {\bf 177}, Nos. 1 \& 2 (1989) 1, North-Holland,
Amsterdam.}

\lref\confrev{P. Ginsparg and J. L. Gardy in {\it Fields, Strings, and
Critical Phenomena}, 1988 Les Houches School, E. Brezin and J. Zinn-Justin,
eds, Elsevier Science Publ., Amsterdam (1989). \hb
J. Bagger, {\it Basic Conformal Field Theory},
Lectures given at 1988 Banff Summer Inst. on Particle and Fields,
Banff, Canada, Aug. 14-27, 1988, HUTP-89/A006, January 1989. }

\lref\CHAN{S. Chandrasekhar, {\it The Mathematical Theory of Black Holes},
Oxford University Press, 1983.}

\lref\KOULU{C. Kounnas and D. L\"ust, Phys. Lett. {\bf 289B} (1992) 56.}

\lref\PERRY{M. J. Perry and E. Teo, Phys. Rev. Lett. {\bf 70} (1993) 2669.\hb
P. Yi, Phys. Rev. {\bf D48} (1993) 2777.}

\lref\GiKi{A. Giveon and E. Kiritsis, Nucl. Phys. {\bf B411} (1994) 487.}

\lref\kar{S.K. Kar and A. Kumar, Phys. Lett. {\bf 291B} (1992) 246.}

\lref\NW{C. Nappi and E. Witten, Phys. Rev. Lett. {\bf 71} (1993) 3751.}

\lref\HK{M. B. Halpern and E. Kiritsis,
Mod. Phys. Lett. {\bf A4} (1989) 1373.}

\lref\MOR{A.Yu. Morozov, A.M. Perelomov, A.A. Rosly, M.A. Shifman and
A.V. Turbiner, Int. J. Mod. Phys. {\bf A5} (1990) 803.}

\lref\KK{E. Kiritsis and C. Kounnas, Phys. Lett. {\bf B320} (1994) 264.}

\lref\KST{K. Sfetsos and A.A. Tseytlin, Phys. Rev. {\bf D49} (1994) 2933.}

\lref\KSTh{K. Sfetsos and A.A. Tseytlin, Nucl. Phys. {\bf B415} (1994) 116.}

\lref\KP{S.P. Khastgir and A. Kumar, ``Singular limits and string solutions'',
IP/BBSR/93-72, hepth/9311048.}

\lref\etc{K. Sfetsos, Phys. Lett. {\bf B324} (1994) 335.}

\lref\KTone{C. Klimcik and A.A. Tseytlin, Phys. Lett. {\bf B323} (1994) 305.}

\lref\KTtwo{C. Klimcik and A.A. Tseytlin, ``Exact four dimensional string
solutions and Toda-like sigma models from `null-gauged' WZNW theories,
Imperial/TP/93-94/17, PRA-HEP 94/1, hepth/9402120.}

\lref\saletan{E.J. Saletan, J. Math. Phys. {\bf 2} (1961) 1.}
\lref\jao{D. Cangemi and R. Jackiw, Phys. Rev. Lett. {\bf 69} (1992) 233.}
\lref\jat{D. Cangemi and R. Jackiw, Ann. Phys. (NY) {\bf 225} (1993) 229.}

\lref\ORS{ D. I. Olive, E. Rabinovici and A. Schwimmer, Phys. Lett. {\bf B321}
(1994) 361.}

\lref\edc{K. Sfetsos, ``Exact String Backgrounds from WZW models based on
Non-semi-simple groups'', THU-93/31, hepth/9311093, to appear in
Int. J. Mod. Phys. {\bf A} (1994).}

\lref\sfedual{K. Sfetsos, Phys. Rev. {\bf D50} (1994) 2784.}

\lref\grnonab{A. Giveon and M. Rocek, Nucl. Phys. {\bf B421} (1994) 173.}

\lref\wigner{E. ${\rm In\ddot on\ddot u}$ and E.P. Wigner,
Proc. Natl. Acad. Sci. U. S. {\bf 39} (1953) 510.}

\lref\HY{ J. Yamron and M.B. Halpern, Nucl. Phys. {\bf B351} (1991) 333.}
\lref\HB{K. Bardakci and M.B. Halpern, Phys. Rev. {\bf D3} (1971) 2493.}
\lref\MBH{M.B. Halpern, Phys. Rev {\bf D4} (1971) 2398.}
\lref\balog{J. Balog, L. O'Raifeartaigh, P. Forgacs and A. Wipf,
Nucl. Phys. {\bf B325} (1989) 225.}
\lref\kacm{V. G. Kac, Funct. Appl. {\bf 1} (1967) 328.\hb
R.V. Moody, Bull. Am. Math. Soc. {\bf 73} (1967) 217.}

\lref\TSEma{A.A. Tseytlin, Nucl. Phys. {\bf B418} (1994) 173.}

\lref\BCHma{J. de Boer, K. Clubock and M.B. Halpern, ``Linearized form
of the Generic Affine-Virasoro Action'', UCB-PTH-93/34, hepth/9312094.}

\lref\AABL{E. Alvarez, L.Alvarez-Gaume, J.L.F. Barbon and Y. Lozano,
Nucl. Phys. {\bf B415} (1994) 71.}

\lref\noumo{N. Mohammedi,  Phys. Lett. {\bf B325} (1994) 371.}

\lref\sfetse{K. Sfetsos and A.A. Tseytlin, unpublished (February 1994).}
\lref\figsonia{J.M. Figueroa-O'Farrill and S. Stanciu, Phys. Lett. {\bf B327}
(\1994) 40.}

\lref\ALLleadord{E. Witten, Phys. Rev. {\bf D44} (1991) 314.\hb
J.B. Horne and G.T. Horowitz, Nucl. Phys. {\bf B368} (1992) 444.\hb
M. Crescimanno, Mod. Phys. Lett. {\bf A7} (1992) 489.\hb
I. Bars and K. Sfetsos, Mod. Phys. Lett. {\bf A7} (1992) 1091.\hb
E.S. Fradkin and V.Ya. Linetsky, Phys. Lett. {\bf 277B} (1992) 73. \hb
I. Bars and K. Sfetsos, Phys. Lett. {\bf 277B} (1992) 269.\hb
P. Horava, Phys. Lett. {\bf 278B} (1992) 101.\hb
E. Raiten, ``Perturbations of a Stringy Black Hole'', Fermilab-Pub 91-338-T.\hb
P. Ginsparg and F. Quevedo,  Nucl. Phys. {\bf B385} (1992) 527.\hb
more}

\lref\ALLexact{
R. Dijkgraaf, E. Verlinde and H. Verlinde, Nucl. Phys. {\bf B371} (1992) 269.
\hb
I. Bars and K. Sfetsos, Phys. Rev. {\bf D46} (1992) 4510.;
Phys. Lett. {\bf 301B} (1993) 183.\hb
K. Sfetsos, Nucl. Phys. {\bf B389} (1993) 424.\hb
Gerhon}

\lref\DHooft{T. Dray and G. 't Hooft, Nucl. Phys. {\bf B253} (1985)
173.}
\lref\MyPe{ R.C. Myers and M.J. Perry, Ann. of Phys. {\bf 172} (1986)
304.}
\lref\GiMa{G. Gibbons and K. Maeda, Nucl. Phys. {\bf B298} (1988) 741.}
\lref\KLOPV{R. Kallosh, A. Linde, T. Ort\'\i n, A. Peet and A. Van Proeyen,
Phys. Rev. {\bf D46} (1992) 5278.}

\lref\lousan{ C.O. Loust\'o and N. S\'anchez, Int. J. Mod. Phys. {\bf A5}
(1990) 915; Nucl. Phys. {\bf B355} (1991) 231.\hb
V. Ferrari and P. Pendenza, Gen. Rel. Grav. {\bf 22} (1990)
1105.\hb
H. Balasin and H. Nachbagauer, Class. Quantum Grav. {\bf 12} (1995) 707. \hb
K. Hayashi and T. Samura, Phys. Rev. {\bf D50} (1994) 3666. \hb
S. Das and P. Majumdar, Phys. Lett. {\bf B348} (1995) 349.}


\lref\lousanII{ C.O. Loust\'o and N. S\'anchez, ``Scattering processes at the
Planck scale'', UAB-FT-353, gr-qc/9410041 and references therein.}

\lref\DHooftII{ T. Dray and G. 't Hooft, Commun. Math. Phys. {\bf 99}
(1985) 613.}
\lref\DHooftIII{T. Dray and G. 't Hooft,
Class. Quant. Grav. {\bf 3} (1986) 825.}

\lref\DHooftII{ T. Dray and G. 't Hooft, Commun. Math. Phys. {\bf 99}
(1985) 613; Class. Quant. Grav. {\bf 3} (1986) 825.}

\lref\VV{H. Verlinde and E. Verlinde, Nucl. Phys. {\bf B371} (1992)
246.}

\lref\aisexl{P.C. Aichelburg and R.U. Sexl, Gen. Rel. Grav. {\bf 2}
(1971) 303.}
\lref\cagan{C.G. Callan and Z. Gan, Nucl. Phys. {\bf B272} (1986) 647.}
\lref\hooft{ G. 't Hooft, Nucl. Phys. {\bf B335} (1990) 138.}
\lref\hoo{G. 't Hooft, Phys. Lett. {\bf B198} (1987) 61;
Nucl. Phys. {\bf B304} (1988) 867.}

\lref\tasepr{ I.S. Gradshteyn and I.M. Ryznik, {\it Tables of integrals,
Series and Products}, Academic, New York, (1980). }
\lref\HOTA{ M. Hotta and M. Tanaka, Class. Quantum Grav. {\bf 10} (1993) 307.}
\lref\otth{ V. Ferrari and P. Pendenza, Gen. Rel. Grav. {\bf 22} (1990)
1105.\hb
H. Balasin and H. Nachbagauer, Class. Quantum Grav. {\bf 12} (1995) 707. \hb
K. Hayashi and T. Samura, Phys. Rev. {\bf D50} (1994) 3666.}

\lref\KSHM{ D. Kramer, H. Stephani, E. Herlt and M. MacCallum,
{\it Exact Solutions of Einstein's Field Equations}, Cambridge (1980).}

\lref\lathos{ C.O. Loust\'o and N. S\'anchez, Phys. Lett. {\bf B220} (1989)
55.}

\lref\BAIT{ M. Bander and C. Itzykson, Rev. of Mod. Phys. {\bf 38} (1966) 330;
Rev. of Mod. Phys. {\bf 38} (1966) 346.}

\lref\rpen{R. Penrose, in General Relativity: papers in honour of J.L. Synge,
ed. L. O'Raifeartaigh (Clarendon, Oxford, 1972) 101.}
\lref\AGV{L.N. Lipatov, Nucl. Phys. {\bf B365} (1991) 614.\hb
R. Jackiw, D. Kabat and M. Ortiz,
Phys. Lett. {\bf B277} (1992) 148.\hb
D. Kabat and M. Ortiz, Nucl. Phys. {\bf B388} (1992) 570.\hb
D. Amati, M. Ciafaloni and G. Veneziano,
Nucl. Phys. {\bf B403} (1993) 707.}


\lref\KPen{ K.A. Khan and R. Penrose, Nature (London) {\bf 229} (1971) 185.}
\lref\DOVE{M. Dorca and E. Verdaguer, Nucl. Phys. {\bf B403} (1993) 770.}

\lref\xanth{S. Cahndrasekhar and B. Xanthopoulos }

\lref\stplane{K. Sfetsos and A.A. Tseytlin, Nucl. Phys. {\bf B427} (1994) 245.}

\lref\shock{ K. Sfetsos, Nucl. Phys. {\bf B436} (1995) 721.}

\lref\gardiner{C.W. Gardiner, {\it Handbook of stochastic methods for Physics,
Chemistry and the Natural sciences}, Berlin, Springer, 1983.}

\lref\CHSW{ S. Chaudhuri and J.A. Schwarz, Phys. Lett. {\bf B219} (1989) 291.}

\lref\DJHOT{ D.H. Hartley, M. \"Onder and R.W. Tucker,
Class. Quantum Grav. {\bf 6} (1989) 1301\hb
T. Dray and P. Joshi, Class. Quantum Grav. {\bf 7} (1990) 41.}

\lref\CHS{ C.G. Callan, J.A. Harvey and A. Strominger, Nucl. Phys. {\bf B359}
(1991) 611.}

\lref\schtalk{ J.H. Schwarz,
``Evidence for Non-perturbative String Symmetries'',
CALT-68-1965, hepth/9411178.}

\lref\bakasII{I. Bakas, Phys. Lett. {\bf B343} (1995) 103.}


\hfill {THU-94/17}
\vskip -.3 true cm
\rightline{December 1994}
\vskip -.3 true cm
\rightline {hep-th/9412065}

\bs\bs\bs

\centerline  {\bf STOCHASTIC TACHYON FLUCTUATIONS, MARGINAL DEFORMATIONS }
\centerline {\bf AND SHOCK WAVES IN STRING THEORY }

\vskip 1.00 true cm

\centerline  {  {\bf Konstadinos Sfetsos}{\footnote{$^*$}
 {e-mail address: sfetsos@fys.ruu.nl }}                                     }

\bigskip

\centerline {Institute for Theoretical Physics }
\centerline {Utrecht University}
\centerline {Princetonplein 5, TA 3508}
\centerline{ The Netherlands }


\vskip 1.30 true cm

\centerline{ABSTRACT}

Starting with exact solutions to string theory on curved spacetimes
we obtain deformations that represent gravitational shock waves.
These may exist in the presence or absence of sources.
Sources are effectively induced by a tachyon field that randomly fluctuates
around a zero condensate value.
It is shown that at the level of the underlying conformal field theory (CFT)
these deformations are marginal and moreover all $\a'$-corrections are
taken into account.
Explicit results are given when the original undeformed
4-dimensional backgrounds
correspond to tensor products of combinations of 2-dimensional CFT's,
for instance $SL(2,\IR)/\IR \otimes SU(2)/U(1)$.

\vskip .3 true cm

\vfill\eject


\newsec{ Introduction }

Gravitational shock waves in general relativity have been considered
in depth in the past as well as more recently, with the
prototype example being the
shock wave due to a massless particle moving in a flat \MI\ background
\aisexl.
The generalization to the case where the particle moves
along a null hypersurface of a more general class of vacuum solutions to
Einstein's equations was found in \DHooft\ and for the cases where there are
non-trivial matter fields and a cosmological constant in \shock.
Explicit results were given when the curved background geometry is the \SW\
black hole in \DHooft, and for the cases of the \RN\ charged black hole,
the De-Sitter space, and the \SW-de-Sitter black hole in \shock.
Other interesting solutions representing the gravitational
field of massless particles with extra quantum numbers (charge, spin),
cosmic strings, or monopoles in a flat \MI\ background
\refs{\lousan}, or in De-Sitter space \HOTA,
have been obtained by infinitely boosting
\refs{\aisexl,\DHooft} known solutions representing curved spacetimes.
For the cases where instead of a massless particle there is
a distribution of massless matter, such as spherical and planar shells, see
\refs{\DHooftII,\DJHOT}.

The main motivation for dealing with gravitational shock waves is that, as
was argued in \hooft, gravitational interactions
dominate any other type of interaction at Planckian energies
(see \refs{\hoo,\VV,\AGV,\lousanII}) and that in an $S$-matrix approach to
black hole
physics \hooft, one needs to take into account the interactions between
Hawking emitted and infalling particles as well as their effect on the
original black hole geometry.
Thus, having the exact solutions to Einstein's equations
(and for that matter to any other theory of gravity)
of a background geometry coupled to a distribution of massless matter
moving along a null hypersurface is equivalent to fully
taking into account all classical backreaction-type effects.

The purpose of this paper is to analyze gravitational shock waves in the
context of string theory.
This was partially done in \shock, but from a general
relativity point of view. However, as we shall see,
the origin of such solutions in string
theory is different from that in general relativity.
Moreover, new features
will be found and a direct connection with the underlying conformal field
theory (CFT) will be made.
The paper is organized as follows: In section 2 we develop the
necessary formalism and obtain the general condition for being
able to introduce a shock wave in a quite general class of solutions to
string theory with two different methods.
One is the general relativity inspired traditional method \DHooft,
where one essentially solves the $\b$-function equations assuming
that they are satisfied by the background geometry fields.
The second method, which is new and uses CFT techniques,
reveals that the shock waves
correspond to marginal perturbations of the CFT corresponding to the original
background. It yields the same results as the more traditional method in a
straightforward way requiring however much less effort.
In addition, as we shall see, it is applicable to more general situations.
We also show that random fluctuations of the tachyon field around its zero
average value
effectively produce source terms for gravitational shock waves, which
nevertheless may exist even in the absence of sources.
In section 3 we apply the general formalism to several cases
where the background
fields correspond to tensor products of various combinations of
2-dimensional exact CFT's. We end the paper with concluding remarks and
a discussion in section 4.
In order to facilitate the computations of section 2 we have written appendix
A containing components of various useful tensors and
appendix C containing elements of stochastic calculus. In Appendix B we use
the CFT method to find shock waves on more general backgrounds
than the ones considered in section 2.

\newsec{ General formalism and results }

Consider the string background in $d$ spacetime dimensions
that comprises a metric, an antisymmetric tensor, and a dilaton field
given by
\eqn\metrd{\eqalign{
& ds^2= 2\ A(u,v)\ du dv\ +\ g(u,v)\ h_{ij} (x)\ dx^i dx^j \cr
& B= 2 B_{ui}(u,v,x)\ du\wedge dx^i + B_{ij}(u,v,x)\ dx^i \wedge dx^j \cr
& \Phi=\Phi(u,v,x)\ ,\cr} }
with $ (i,j=1,2,\dots , d-2) $. Let us suppose that a `disturbance' is
introduced (whose origin and nature will be examined later in this section)
with the net effect that the spacetime is described by \metrd\ only
for $u<0$, whereas for $u>0$ we should replace in \metrd\ $v\to v+ f(x)$
and $dv\to dv + \fpi dx^i$. Thus the two spacetimes for $u<0$ and $u>0$ are
glued together along the null hypersurface $u=0$ \DHooft.
A compact way to represent the spacetime fields is by using the Heaviside
step function $\vartheta = \vartheta(u)$
\eqn\metrgl{\eqalign{
& ds^2=  2\ A(u,v+\vartheta f )\ du (dv + \vartheta \fpi dx^i)
\ +\ g(u,v+\vartheta f)\ h_{ij} (x)\ dx^i dx^j \cr
& B= 2 B_{ui}(u,v+\vartheta f,x)\ du\wedge dx^i +
B_{ij}(u,v+\vartheta f,x)\ dx^i \wedge dx^j \cr
&\Phi=\Phi(u,v+\vartheta f,x)\ .\cr }}
The coordinate change
\eqn\cooch{ u\to u\ ,\quad v\to v- f(x) \vartheta(u) \ ,\quad x^i\to x^i \ ,}
gives a form where various tensors are easier to compute
\eqn\metrgll{\eqalign{ & ds^2=  2\ A(u,v)\ du dv
+ g(u,v)\ h_{ij} (x)\ dx^i dx^j + F(u,v,x)\ du^2 \cr
& B= 2 B_{ui}(u,v,x)\ du\wedge dx^i + B_{ij} (u,v,x)\ dx^i \wedge dx^j \cr
& \Phi=\Phi(u,v,x)\ ,\cr }}
where $F(u,v,x)\equiv -2 A(u,v) f(x) \d(u)$, and $\d(u)={d \vartheta(u)\ov du}$
is a $\d$-function. In order to determine
the shift function $f(x)$ we require that the $\b$-function equations
that govern the dynamics of the lowest modes of the string are satisfied\foot{
For zero tachyon field ($T=0$) these are the one loop $\b$-function equations
corresponding to the metric, antisymmetric tensor, and dilaton fields
(see, for instance, \refs{\FRATSE,\CALLAN}).
The tachyonic contributions \DASA\ are non-perturbative in the loop expansion.}
\eqn\streinst{\eqalign{
& R_{\m\n} - D_{\m} D_{\n} \Phi
- {1\ov 4} H_{\m\r\l} H_{\n}{}^{\r\l} = \partial_{\m} T \partial_{\n} T \cr
& D_{\l} (e^{\Phi} H^{\l}{}_{\m\n})=0 \cr
& C= d + {3\ov 2}\a' \bl( D^2\Phi +(D\Phi)^2 - {1\ov 6} H^2_{\m\n\l}
-(DT)^2 - 2 V(T) \br) \cr
& D^2 T + D_{\m} \Phi D^{\m} T = V'(T) \ ,\cr} }
where the tachyon potential is $V(T)\sim T^2$ and $C$ denotes the central
charge. By assumption, in the bulk
of the space ($u\neq 0$) and with zero tachyon field, i.e., $T=0$,
these equations are automatically satisfied by the
background fields \metrgll\ or equivalently \metrd.
However, as one might expect, there are extra contributions from the
boundary at $u=0$ (in fact multiplied by a $\d(u)$-function).
Using the results of appendix A we find that consistency requires
the conditions
\eqn\conff{ \Apv = \gpv= \Phi_{,v} = 0 \quad {\rm at}\ u=0\ .}
In addition the shift function $f(x)$ is obtained by solving the linear
differential equation\foot{
In order to cast it into that form we have used the fact that
\eqn\conduv{ {A_{,uv}\ov A} + {d-2\ov 2} {g_{,uv}\ov g} + \Phi_{,uv}
-{1\ov 2 g A}H_{uvi}H_{uvj} h^{ij}  =0
\quad {\rm at }\ u=0 \ .}
This is nothing but the $(u,v)$-component of the metric $\b$-function computed
at $u=0$ and simplified by using \conff\ and the fact that $H_{vij}=0$ at $u=0$
(this follows from the $(v,v)$ and $(v,i)$-component of the metric
$\b$-function).}
\eqn\condit{\eqalign{
& \triangle f(x)  -  c(x) f(x)  =  2\pi b\ \d^{(d-2)}(x-x') \cr
& c(x)\equiv {1\ov A} \bl({d-2\ov 2} g_{,uv} +  g \Phi_{,uv}\br)\ ,
\quad   b=k {g\ov A} \ ,\cr }}
where all functions are computed at $u=0$ and the Laplacian is defined as
\eqn\laplaa{ \triangle = {1\ov e^{\Phi} \sqrt{ h}} \partial_i e^{\Phi}\sqrt{ h}
\ h^{ij} \partial_j \ .}
The conditions \conff\condit\ were derived by examining
the metric $\b$-function. The rest of the equations give no additional
information.\foot{ Had we included a $B_{vi}$
component in the antisymmetric tensor we should have required that,
in addition to \conff\condit, $B_{vi}=O(u^n)$ and $B_{vi,v}=O(u^m)$,
with $n>\ha$ and $m>1$, near $u=0$.
If $B_{vi,v}=O(u)$ an additional
non-linear term $O(f^3)$ seems to appear in \condit.
Since this could be a new feature it might be interesting to explore it
further.}
We have also included a specific source term (with strength proportional to
a constant $k$) containing a $\d$-function defined in the transverse
$x^i$-space and normalized with
respect to the `string measure' $e^{\Phi}\sqrt{h}$ computed at $u=0$, namely,
\eqn\noorm{ \int_{M_\perp} d^{d-2}x\ e^{\Phi} \sqrt{h}\ \d^{(d-2)}(x)=1\ ,
\quad {\rm at}\ u=0\ .}
The conditions \conff\ and \condit\ are the string theory analogue of the
similar conditions found in the context of Einstein's general relativity
in $d$-dimensions \shock\ and they reduce to them for a constant dilaton field.

So far we have given no explanation at all about
the origin of the source term present on the right hand
side of the equation in \condit. Strictly speaking for a zero tachyon field it
should be zero. In fact in certain cases (but not in general)
that differential equation with the zero right hand side has a solution.
It can be shown (see also \shock) that in such cases the term
\eqn\pertt{ -2 \int d^2z\ \d(u) f(x) A(u,v)\ \del u\bd u \ ,}
corresponds to a marginal perturbation of the original 2-dimensional $\s$-model
action for the background \metrd\ in the sense that it
solves the corresponding conditions as they were found, to leading order in
$\a'$, in \cagan\ (in this paper it was assumed that the antisymmetric tensor
was identically zero, but presumably a generalization to the non-zero
case exists). In fact it can be shown that these
conditions reduce, in our case, to just
\conff\condit.
Before we explain the origin of the $\d$-function source term on the right hand
side of \condit, let us rederive \conff\condit\ using standard CFT techniques.
We will show that \conff\condit\ are the necessary and sufficient
conditions for \pertt\ to be a marginal perturbation (in fact we will argue
that is is exactly marginal) and that
these conditions hold beyond the one loop approximation,
i.e., are in fact exact to all orders in $\a'$.
This method is considerably faster and could be easily adopted to other
similar situations (see appendix B).
The first step is to show that $\del u$ has conformal dimension $(1,0)$
with respect to the energy momentum tensor corresponding to the background
\metrd\ in the limit $u\to 0$ (remember the $\d(u)$-function).
The holomorphic component of the energy momentum tensor is
\eqn\enee{ T= {1\ov \a'} \bl( :A \del u \del v: + :g h_{ij} \del x^i \del x^j:
\br)  + :\del^2 \Phi: \ ,}
where a proper regularization prescription is implied. In general finding
the operator product expansions (OPE's) for the fields $u,v,x^i$ is very
difficult due to non-linearities. However, close to $u=0$ we can infer that
\eqn\ooopp{ u(z,\z) v(w,\bar w) = {1\ov A} \ln |z-w|^2 + \dots \ ,}
where the ellipsis denotes terms that vanish as $z\to w$ (and $\z\to \bar w$).
Therefore $\del u$ at $u=0$ will be a dimension $(1,0)$ operator with respect
to \enee\ and its antiholomorphic partner,
provided that all possible anomalies arising
from contractions with the fields $u$, $v$ in $A(u,v)$, $g(u,v)$ and
$\Phi(u,v)$ vanish.
It is easily seen that conditions \conff\ guarantee exactly that.
Rephrasing, conditions \conff\ guarantee that close to $u=0$ the CFT for
the longitudinal part is effectively that of two free bosons.
Analogously $\bd u$ at $u=0$ has dimension $(0,1)$
and thus the operator $\del u \bd u$ has dimension $(1,1)$ at $u=0$.
Having established that, we need to discover the condition $A(u,v) f(x) \d(u)$
has to satisfy in order to really be a function,
i.e., have dimension $(0,0)$. Then the term \pertt\ will
correspond to a marginal perturbation (but not in principle exactly marginal).
On general grounds, the dimension $D$ of $A(u,v) f(x)\d(u)$ is determined from
the eigenvalue, Klein-Gordon type, equation
\eqn\dimso{ -{1\ov e^{\Phi} \sqrt{-G}} \del_{\m} e^{\Phi} \sqrt{-G} G^{\m\n}
\del_{\n}\ A(u,v) f(x)\d(u) = D\ A(u,v) f(x) \d(u) \ ,}
where $\sqrt{-G}= A g^{{d-2\ov 2}} \sqrt{h}$. Demanding that $D=0$ and
after simplifying using \conff, the above equation reduces exactly to \condit\
with a zero right hand side.

So far we have set the tachyon field to zero.
However, we can slightly relax this condition by demanding
that only its average value is zero but otherwise it can randomly fluctuate.
The origin of such fluctuations can be either statistical, due to our inability
to determine its precise value, or of more fundamental nature, as remnants of
stringy phenomena at high energies
(ambiguity in the usual spacetime description at Planck scale, etc.)
that effectively make, at low energies, the tachyon to appear fluctuating,
or a combination of both.
We will show that these fluctuations induce (upon taking the average of the
tachyon energy momentum tensor)
the non-zero source term on the right hand side of \condit.
Specifically let us consider a tachyon $T$ that factorizes as
\eqn\Ttf{ T(u,x)= \tau(u)\ h(u,x)\ ,}
where $\tau(u)$ is given by the stochastic integral in the Ito calculus,
\eqn\stocc{\tau(u) =
\sqrt{ 2 \ov  \pi} \int_{-\infty}^{\infty} dw(t)\ e^{it u} \ ,}
and $h(u,x)$ is any deterministic function that behaves as
\eqn\dethux{ h(u,x) = \sqrt{\r(x)}\ u + O(u^2)\ ,}
where $\r(x)$ is a density-like function.
Elements of stochastic calculus are given in appendix C.
Using them we compute the expectation values
\eqn\exptau{ \eqalign{ & \langle \tau^2(u) \rangle = 2\d(u) \cr
&\langle \tau(u) \tau'(u) \rangle = \d'(u) \cr
& \langle \tau'^2(u) \rangle = \ha \d''(u) \ ,\cr }}
where the prime denotes differentiation with respect to $u$.
{}From these and the leading order behavior of $h(u,x)$ near
$u=0$ we obtain expectation values involving directly the tachyon field
\eqn\exptaa { \eqalign { & \langle T \rangle = 0 \cr
& \langle \partial_i T \partial_j T \rangle = 2\ \partial_i\sqrt{\r(x)}
\partial_j \sqrt{\r(x)}\ u^2 \d(u) =0 \cr
& \langle \partial_u
T \partial_i T \rangle = \ha \partial_i {\r(x)}
( u^2 \d'(u) + 2 u \d(u) ) = 0 \cr
& \langle (\partial_u T)^2 \rangle =
\r(x) \bl( 2 \d(u) + 2 u\d'(u) +\ha u^2 \d''(u)\br) = \r(x)\d(u) \ .\cr }}
It is understood that \exptaa\ hold in a distribution sense with
respect to the variable $u$ and therefore integration over a smooth function
of $u$ is implied.

The upshot of this analysis is that by taking the expectation value of the
$\b$-function equations\foot{ Notice that without taking the average the
$\b$-functions are not satisfied due to the tachyon stochastic fluctuations.
However, our philosophy is that they need only be satisfied in the `average'.}
and by using \exptaa\ we obtain the same equations we
would have obtained had we set the tachyon field to zero {\it except} from
a source term on the right hand side of the $\m=\n=u$ component of the metric
$\b$-function. Namely
\eqn\souu{ R_{uu}-D_uD_u \Phi -{1\ov 4} (H^2)_{uu} = \r(x) \d(u)\ ,}
thus proving that random fluctuations around a zero tachyon background
induce source
terms for gravitational shock waves in string theory.\foot{Tachyon
fluctuations are not the only possible such source. For instance, if $A(u,v)$,
$g(u,v)$ are constant functions and $\Phi(u,v,x)=\Phi(x) + k u \vartheta(u)$
we obtain a source term with uniform distribution $\r(x)=k=\const$,
because in this case $D_uD_u\Phi= k \d(u)$.}
Since we are interested in the Green's functions we chose
$\r(x)=k\ \d^{(d-2)}(x-x')$. The result for any other distribution $\r(x)$
is given by the integral $\int [dx'] f(x,x') \r(x')$, where the measure
$[dx']$ is defined according to \noorm.
Let us also mention that the central charge coincides with the value obtained
by simply having a vanishing tachyon.
Not only do the tachyon dependent terms vanish
upon taking the average but also the derivatives of them with respect to all
fields (including $u$) as well.
Also it is obvious that in this case the term \pertt\ does not represent a
marginal perturbation by itself.
However, its conformal anomaly balances that of the stochastically fluctuating
tachyon. This is precisely the meaning of \condit.

In the CFT approach to deriving \conff\condit\ the backgrounds \metrd\metrgll\
are supposed to satisfy the $\b$-function equations to all
orders in perturbation theory in powers of $\a'$ in the standard `conformal
scheme' (see \KST), where also the tachyon equation takes the simple form given
by the last equation in \streinst. Thus, we conclude that \conff\condit\ are
indeed valid to all orders in conformal perturbation theory in the standard
`conformal scheme'.

The final comment is on
whether or not \pertt\ represents an exactly marginal perturbation.
This would be the case if we can argue that higher order terms in $f$ do not
spoil conformal invariance.
In fact such non-linearity in $f$ terms does appear when we consider
the $\b$-function equations (see (A.2),(A.3)).
However, in that case one can show that these terms vanish in a distribution
sense \shock. In the present case we can argue that anomaly terms
proportional to powers of $f$ in the Virasoro
algebra generated by the energy momentum tensor corresponding to the
background \metrgll\ also vanish, as follows. A possible such term contains
the factor
\eqn\anfa{
\prod_i {\partial^{n_i+n'_i} A\ov \partial v^{n_i} \partial {u^{n'_i}}}\
{\partial^{m_i+m'_i} g\ov \partial v^{m_i} \partial {u^{m'_i}}}\
{\partial^{l_i+l'_i} \Phi\ov \partial v^{l_i} \partial {u^{l'_i}}}
\ f^n \d^n \ ,}
where the relation between the various integers
\eqn\uiuh{\sum_i ( n_i +m_i +l_i ) = 2 n + \sum_i ( n'_i +m'_i +l'_i) \geq 2n }
should also hold. This follows
from the facts that near $u=0$ there is an invariance of the theory described
by \metrd\ under $u,v$ interchange (cf.(A.1)-(A.4)) and that the
energy momentum tensor must be invariant under this symmetry.
Since for regular functions around $u=0$ each derivative with respect to
$v$ contributes a power of $u$, it can easily be seen
that \anfa\ vanishes as a distribution thanks to the inequality in \uiuh.
Let us once more emphasize that the term \pertt\ corresponding to an
exactly marginal perturbation is a consequence of \conff\condit\ and the
presence of the $\d(u)$-function, and that this is not generally true for
marginal perturbations with abelian chiral currents \CHSW.

\newsec {Applications}

\def\fpa{\phi_{\parallel}}
\def\fpe{\phi_{\perp}}

In all of our applications we start with the direct product of two
2-dimensional CFT's with a metric and dilaton of the form
(the antisymmetric tensor is zero)
\eqn\apform{\eqalign{ & ds^2 = 2 A(u,v)\ dudv +  h_{ij}(x)\ dx^i dx^j \cr
& \Phi(u,v,x) = \fpa(u,v) + \fpe(x^1,x^2) \ ,\cr }}
namely, the longitudinal and the transverse parts are decoupled.
The longitudinal CFT provides the `time' coordinate for the metric of
our model and
we will take it to be either the one corresponding to the coset
$SL(2,\IR)_{-k}/\IR$ \refs{\BN,\WIT}, or that corresponding to the flat
2-dimensional \MI\ space.\foot{ It can be shown that if we take as the
longitudinal CFT the one corresponding to the dual of the 2-dimensional
\MI\ space (corresponding
to the coset $E^c_2/(\IR\otimes U(1))$ \refs{\etc,\stplane})
the conditions \conff\ are
not satisfied.}
As the transverse part we will take either the compact
coset $SU(2)_k/U(1)$ \BCR\ or flat 2-dimensional space (with possibly a linear
dilaton) or the dual to the 2-dimensional flat space.
The coupling between the two CFT's is only due to the term \pertt\
corresponding to the shift function $f(x)$
which satisfies the differential equation \condit, with constant $c$.
The solution to this differential equation can be expressed as an infinite sum
over eigenfunctions of the Laplacian \laplaa. The result is easily found to be
(we ignore possible solutions of the homogeneous equation)
\eqn\ressuu{ f(x;x') = -2\pi b \sum_N { \Psi_N(x) \Phi_N^*(x') \ov E_N+ c} \ ,}
where we denoted by $N$ all possible quantum numbers arising from the
eigenvalue equation
\eqn\eigee{ \triangle \Psi_N(x) = - E_N \Psi_N(x) \ .}
Notice that since we are dealing with compact manifolds corresponding to the
transverse metric the Laplacian \laplaa\ is a negative definite operator.
For this reason the minus sign on the right hand side of \eigee\ implies
that $E_N\ge 0$.
The eigenfunctions satisfy the orthonormalization condition
\eqn\orthoo{ \int_{\Sigma} d^2 x e^{\fpe} \sqrt{h} \Psi_N(x) \Phi_M^*(x)
= \d_{N,M} \ ,}
and the completeness relation
\eqn\coomp{ \sum_N \Phi_N(x) \Phi_N^*(x') = \d^{(2)}(x-x') \ .}
Notice that, in accordance with \laplaa\ and \noorm\
the `string measure' $e^{\fpe} \sqrt{h}$ has been used and not
just $\sqrt{h}$. Also that, in the sourceless case,
\eigee\ is exactly of the form \condit.
Therefore, if $c<0$ and moreover coincides with one of the
eigenvalues, i.e., $c=-E_N$ for some $N$, the corresponding eigenfunction
$\Psi_N$ gives the solution for the shift function $f$.
In the case with source, solution \ressuu\ is not valid if $c$ coincides with
any of the eigenvalues $E_N$. Then the solution will be given in terms of
the `partner' of the $\Psi_N$ in \eigee\ which has the appropriate
singular short distance behaviour that produces a $\d$-function
(For specific examples see \shock. An important one is gravitational
shock waves with sources in 4-dimensional De-Sitter space).
These cases will not be considered in this paper.

Let us also mention that for notation, conventions, and
various results involving special functions we will use \tasepr.

\subsec { $SL(2,\IR)_{-k}/\IR \otimes \IR \otimes \IR'$ }

For this model the metric and dilaton are
\eqn\sllrr{\eqalign{& ds^2 = {2 \e\ov uv-1}\ dudv +
a^2 ( dx_1^2  + dx_2^2 ) \cr
& \Phi=\fpa (u,v) + \fpe(x_1) \ ,\qq \fpa=\ln (1-uv), \ \fpe= 2\a_0 x_1 \cr
& -\infty < u,v,x^1, x^2 < \infty \ .\cr } }
where $a$ is a constant and $\e={\rm sign}(k)$, with $(-k)$ being the central
extension of the $SL(2,\IR)_{-k}$ current algebra.\foot{Even though
\conff\condit\ are exact expressions,
for simplicity of the presentation we have chosen to
work in the small $\a'$ limit
corresponding to a high level ($k>>1$) current algebra. The same remark holds
for the other examples we consider in this section. The exact, in $\a'$,
expressions for $f$, in the standard `conformal scheme' (see \KST),
can be found
using the results of \refs{\DVV,\BSexa}.}
For $\e =1$ the causal structure of the
spacetime is that of a black hole \WIT\
with a singularity at future times $t=u+v$, whereas for $\e=-1$ it has
the cosmological interpretation of an
expanding Universe with no singularity at future times $t=u-v$ (see \KOULU).
We have also allowed for the possibility of a linear
dilaton in the transverse part with strength proportional to the constant
$\a_0$.
In this case the eigenvalue equation is
\eqn\edio{ \partial_{x^i}\partial_{x^i}\Psi +
2\a_0 \partial_{x^1} \Psi = -E \Psi\ ,}
which after substituting $\Psi\to e^{-\a_0 x^1} \Psi$ becomes the eigenvalue
equation for the standard Laplace operator in Euclidean flat space
with $E\to E - a_0^2$.
Therefore the eigenfunctions and eigenvalues
(in a definite angular momentum sector when $\a_0=0$) are
\eqn\qpqp{ \Psi_{k,m}(\r,\phi) = {1\ov \sqrt{2\pi}}\ e^{-\a_0 \r\cos \phi}\
e^{im\phi}\ J_m(k\r) \ ,\quad E_k=k^2 + \a_0^2 \ .}
If we denote by $c\equiv \a_0^2 + \e a^2 $ then the solution for the shift
function is
\eqn\fresoo{ f(\r,\phi;\r',\phi')= -b\ e^{-\a_0 (\r\cos \phi + \r' \cos\phi')}
\sum_{m=-\infty}^{\infty} \int_{-\infty}^{\infty} dk {k\ov k^2+c}
e^{im(\phi-\phi')} J_m(k\r) J_m(k\r')\ ,}
which after using a resummation theorem for Bessel functions and computing
an integral becomes
\eqn\sshf{ f(\r,\phi;\a',\phi') =  b\ e^{-\a_0(\r \cos\phi + \r'\cos \phi')}
\cases{- K_0(\sqrt{c} R)\ , \ & if $ \ \ c >0 $
\cr \noalign{\vskip 6pt}
\ln(R)\ ,\ & if $\ \ c =0 $
\cr \noalign{\vskip 6pt}
{\pi\ov 2} N_0(\sqrt{|c|}R)\ ,\ & if $\ \ c < 0 $\ ,\cr }}
where $R=\sqrt{ \r^2 +\r'^2 -2 \r\r' \cos (\phi-\phi')}$.
Another way to obtain the same result without resorting to the general
prescription \ressuu\ is by noticing that after substituting
$f\to e^{-\a_0 x^1} f$ the equation for $f$ is either the Bessel equation
(if $c<0$) or the modified Bessel one (if $c>0$) and that the special functions
in \sshf\ are the only solutions with the appropriate logarithmic
behavior that produces the $\d$-function.
Let us mention that the case of flat space
in the longitudinal part corresponds to letting $\e=0$ in \sshf\ (but not
in \sllrr) be zero. This is in fact the analogue of the result of \aisexl\
for string theory. Notice however that the presence of the linear dilaton
in the transverse part modifies the solution which now depends explicitly on
$\phi$ and also it vanishes exponentially for large $R$'s instead of growing
logarithmically.


In the sourceless case for $c<0$ ($c>0$) a basis of solutions of \condit\
for $f$ is given by \qpqp\ with the replacement
$J_m(k \r)\to J_m(\sqrt{|c|} \r)$ ($I_m(\sqrt{c}\r)$).
For the case $c=0$ ($a=\a_0$) a solution is $f(x^1,x^2)= e^{-\a_0 x^1}
\bl((x^1)^2-(x^2)^2\br)$.

\subsec { $SL(2,\IR)_{-k}/\IR\ \otimes $ (dual to $2d$ flat space) }

\def\rl{\r_{{}_<}}
\def\rr{\r_{{}_>}}

In this case the metric and dilaton are
\eqn\sllrrr{\eqalign{& ds^2 = {2 \e\ov uv-1}\ dudv +
a^2 ( d\r^2 + {4\ov \r^2} \ d\phi^2 ) \cr
& \Phi=\fpa (u,v) + \fpe(\r) \ ,\qq \fpa=\ln (1-uv), \
\fpe= \ln(\r^2) \cr
& -\infty < u,v < \infty\ ,\quad 0 < \r < \infty\ ,\quad \phi\in [0,2\pi]\
,\cr}}
where $a$ is a constant, $\e={\rm sign}(k)$ and
the physical interpration of \sllrrr\ is similar with that in the previous
example.
The background in \sllrrr\
can be obtained if we write the transverse part of the
background \sllrr\ (with $\a_0=0$)
in terms of polar coordinates and then perform a duality
transformation with respect to $\phi$.
Notice that there is no linear dilaton term (as in \sllrr)
for the transverse part
since that would not be consistent with conformal invariance.
The eigenvalue equation to be solved is
\eqn\eqdio{ {1\ov \r } \partial_{\r} \r \partial_{\r} \Psi + {1\ov 4} \r^2
\partial^2_{\phi} \Psi = -E \Psi\ .}
Changing variables as $\r^2 = \xi$ and substituting
\eqn\eqdioo{ \Psi(\r,\phi) = e^{im\phi} e^{-{1\ov 4} |m| \xi} T(\xi) \ ,\quad
m\in Z }
we see that $F(\xi)$ satisfies the Laguerre equation
\eqn\eqdiooo{ \xi F'' + (1-\ha |m| \xi) F' + {1\ov 4} (E-|m|) F =0 \ .}
This has as solution Laguerre polynomials provided   that $E=|m|(2n+1)$.
Therefore the appropriately normalized eigenfunctions and the corresponding
eigenvalues are
\eqn\eqddiof{\eqalign{& \Psi_{n,m}(\r,\phi)= \sqrt{{|m|\ov 2\pi}}\ e^{im\phi}\
e^{-{1\ov 4} |m| \r^2}\ L_n(\ha |m| \r^2) \cr
& E_{n,m} = |m|(2n+1)\ ,\quad n=0,1,\dots, \ m\in Z-\{0\} \ .\cr }}
Notice that we have excluded the value $m=0$ since
the corresponding eigenfunctions
and eigenvalues become zero. In addition to the discrete part of the
spectrum there is also a continuous one exactly when $m=0$.
One way to see that
is to cast \eqdio\ in the form of a Schr\"odinger equation and read off the
corresponding effective potential, which turns out to be
$V(\r)={1\ov 8}( m^2\r^2 - \r^{-2})$.
On general grounds for $m\neq 0$ there are only the bound states with the
discrete eigenvalues \eqddiof\ whereas for $m=0$ the spectrum is continuous
with\foot{In addition to
directly solving \eqdio\ with $\partial_{\phi}\Psi=0$, there is another
way to obtain it from \eqddiof. Namely, by letting $m\to 0$,
$n= \ha k^2/m \to \infty$, $x={1\ov 4} k^2\r^2$, and using
\eqn\uflj{\lim_{n\to \infty} L_n({x\ov n}) = J_0(2 \sqrt{x})\ .}}
\eqn\specon{ \Psi_k(\r) = {1\ov \sqrt{2 \pi}}\ J_0(k\r)\ ,\qq E_k = k^2 \ .}
Let us point out that had we missed \specon\ we would not be able to write down
the completeness relation \coomp.
The solution for the shift function after we compute an integral is
\eqn\sldua{
f(\r,\phi;\r',\phi') = f_0(\r;\r') -2 b
\sum_{n=0}^{\infty}\sum_{m=1}^{\infty} {m \cos m(\phi-\phi') \ov m(2n+1) + c}
e^{-{1\ov 4} m(\r^2 +\r'^2)} L_n(\ha m \r^2) L_n(\ha m \r'^2) \ ,}
where $c=\e a^2$ and
\eqn\fzz{ f_0(\r;\r') = b\
\cases{- I_0(\sqrt{c} \rl) K_0(\sqrt{c} \rr)\ , \ & if $ \ \ c >0 $
\cr \noalign{\vskip 6pt}
{\pi\ov 2} J_0(\sqrt{|c|} \rl) N_0(\sqrt{|c|}\rr)\ ,\
& if $\ \ c < 0 $\ ,\cr } }
is the $\phi$-independent part of the solution. In the case of $c=0$
(corresponding to taking flat \MI\ space in the longitudinal part)
the solution is found to be
\eqn\czzo{ f(\r,\phi;\r',\phi') = b\ \bl( \ln R - \sum_{m=1}^{\infty}
\cos m(\phi-\phi') I_0({1\ov 4}m \rl^2) K_0({1\ov 4} m \rr^2) \br) \ ,}
where, as usual, $\rl$ ($\rr$) denotes the smaler (larger) of $\r,\r'$.

As an important remark let us mention that although certain backgrounds might
be related via a duality transformation, 
the presence of the term \pertt\ destroys (in general) this relationship.
An example is that although \sllrr\ (with $\a_0=0$) and \sllrrr\ are duality
related, after the addition of the term \pertt\ with $f(x)$ given
correspondingly by \sshf\ and \sldua\ there is no duality transformation that
relates them, because now both backgrounds depend explicitly
on the angle $\phi$ and there is no isometry with respect to which
dualization can be done. An exceptional case is for $\r'=0$,
since then \sshf\ does not depend on $\phi$.
Obviously, after dualization the new $f$ is given by only the
first term in \sldua\ (computed at $\r'=0$), namely the $\phi$-independent
part, corresponding to the angular uniform distribution
$\r(r)={1\ov 2\pi r}\d(r)$
(notice that, $\d^{(2)}(x)= {1\ov r} \d(r) \d(\phi)$).

The solution for the shift function in the sourceless case and in a definite
angular momentum sector is for $m=0$ given: by $f\sim I_0(a \r)$ if $\e=1$ and
by $f\sim J_0( a\r)$ if $\e=-1$. For $m\neq 0$ the solution contains a
confluent hypergeometric function, i.e., $f\sim  \Phi\bl(\ha({c\ov 2} +1),
1,\ha |m| \r^2 \br)\ e^{im\phi} e^{-{1\ov 4} |m| \r^2}$, which becomes
\eqddiof\ if $c$ is one of the eigenvalues of the Laplacian.

\subsec {  $SL(2,\IR)_{-k}/\IR \otimes SU(2)_{k'}/U(1)$ }

For this model the corresponding metric and dilaton are 
\eqn\sllsuu{\eqalign{& ds^2 = {2 \e\ov uv-1}\ dudv +
a^2 \bl( d\th^2 + \tan^2{\th\ov 2}\ d\phi^2 \br) \cr
& \Phi=\fpa (u,v) + \fpe(\th) \ ,\qq \fpa=\ln (1-uv), \
\fpe= \ln \cos^2{\th\ov 2} \cr
& -\infty < u,v < \infty\ ,\quad \phi\in [0,2\pi]\ ,\quad \th\in [0,\pi]\ ,
\cr } }
where $\e={\rm sign}(k)$ and $a^2\equiv {k'\ov k}$ and similar physical
interpretation as before.
After we change variable as $x=\cos \th$ equation \eigee\ becomes
\eqn\eqena{ (1-x^2) \partial^2_x \Phi - 2x \partial_x \Psi
+ {1+x\ov 1-x} \partial^2_{\phi} \Psi = -E \Psi \ .}
Further substitution,
\eqn\eqenaa{ \Psi(x,\phi) = (1-x)^{|m|} e^{im \phi} T(x)\ ,\quad m\in  Z\ , }
shows that $T(x)$ satisfies a Jacobi equation
\eqn\eqenaaa{ (1-x^2) T'' - 2 (|m| + (|m| +1) x) T' + (E- |m|) T = 0\ .}
A complete set of normalizable solutions to it
exists (the so called Jacobi polynomials) provided that
$E=n(n+1)+(2n+1)|m|$, where $n$ is an integer.
Using instead of $n$ the non-negative integer $l=n + |m|$ we eventually get
\eqn\eqenaf{\eqalign{
& \Psi_{l,m}(\th,\phi) =  \sqrt{2l+1\ov {4\pi } }\
e^{i m \phi}\ \sin^{2|m|}{\th\ov 2}\ P^{(2|m|,0)}_{l-|m|} ( \cos \th) \cr
& E_{l,m}=l(l+1) - m^2 \ ,\quad  l=0,1,\dots ,\ m=-l,-l+1,\dots , l \ ,\cr } }
where a compatible with \orthoo\coomp\ normalization factor has also
been included.
Notice that the eigenvalues $E_{l,m}$ are exactly what one expects from the
coset construction for $SU(2)_k/U(1)$ for states at the base of the Virasoro
modules and for high level $k$. The two terms that appear are the eigenvalues
of the quadratic Casimirs for $SU(2)$ and $U(1)$ respectively.
It is now straightforward to write down the solution for the shift function
according to \ressuu
\eqn\soslsu{\eqalign{ & f(\th,\phi;\th',\phi')=
-b\sum_{l=0}^{\infty}{l+\ha \ov l(l+1) + c} P_l(\cos \th) P_l(\cos \th') \cr
& -2 b \sum_{l=0}^{\infty} \sum_{m=1}^l { (l+\ha ) \cos m(\phi-\phi')
\ov l(l+1) -m^2 +c }
\sin^{2m} {\th\ov 2} \sin^{2m} {\th'\ov 2} P^{(2m,0)}_{l-m}(\cos \th)
P^{(2m,0)}_{l-m}(\cos \th')  \ ,\cr }    }
where $c=\e a^2$.

It should be possible to obtain expressions \sshf\ (for $\a_0=0$) and
\sldua\ by taking appropriate limits in \soslsu. The reason is that the
corresponding spacetimes are related via limiting procedures.
Specifically, if $\th =\d \r$, $\phi\to 2 \phi$ (that will effectively
change $m\to m/2$ in \eqenaf), $c\to c/\d^2$ with $\d\to 0$
the background \sllsuu\ becomes that of \sllrr\ with $\a_0=0$.
Naively the shift function as given by \soslsu\ becomes zero. However, if we
treat carefully the contribution coming from the $l=k/\d \to \infty$ values in
the sum (which in this limit becomes an integral over $k$) and use
\eqn\ufpm{ \lim_{l\to \infty} \bl({x\ov 2 l}\br)^{\a}
P_l^{(\a,\b)}(\cos {x\ov l}) = J_{\a}(x) \ ,}
for $l={k\ov \d}$, $x=\r k$ and $\a=m$, $\b=0$ we obtain \fresoo.
Next we let $\th=\pi-\d \r$, $\phi\to \phi \d^2$, $c\to c/\d^2$ with
$\d\to 0$. Because of the rescaling in $\phi$ the new $\phi$ will not
be periodic and the corresponding eigenvalue $m\to m/\d^2$ will not be an
integer. To make $\phi$ again periodic we identify points in the real line,
i.e., we quotient with a discrete subgroup of $\IR$.
Then the background \sllsuu\ becomes that of \sllrrr. Then by letting
$l=m/\d^2 + n$ and using
\eqn\lilli{ \eqalign{& \lim_{\a \to \infty} P_n^{(\a,\b)}(1-{2x\ov \a}) =
(-1)^n L_n^{\b}(x) \cr
& \lim_{\a \to \infty} \cos^{\a}(\sqrt{{x\ov \a}}) = e^{-x/2}\ ,\cr }}
with $x=\ha m\r^2$, $\a=2m/\d^2$, $\b=0$ we obtain from the double sum term of
\soslsu\ a similar term in \sldua.
Obviously the case $m=0$ corresponding to the first term in \soslsu\
requires a different treatment since in this case we cannot take the
$m\to \infty$ limit. In fact
this term in \soslsu\ becomes $f_0(\r;\r')$ in \sldua\ after using \ufpm\ for
$l={k\ov \d}$, $x=\r k$, $\a=\b=0$, replacing the sumation over $l$ with
an integral over the continuous variable $k$ and evaluating this integral.

Let us finally mention that if the transverse CFT is the coset
$SL(2,\IR)_{-k'}/\IR$
corresponding to a `Euclidian black hole' then we should analytically continue
$\th\to i r$ or $\th\to \pi + i r$ and in \soslsu\ sum over
the appropriate representations functions for the non-compact group
$SL(2,\IR)$.

\newsec{ Concluding remarks and discussion }

In this paper we investigated gravitational shock waves in string theory.
We started with quite a general class of background solutions to the one loop
$\b$-function equations and found the conditions (see \conff\condit)
that should be fulfilled in
order to be able to introduce a shock wave via a coordinate shift.
These shock waves may exist with or without sources. In the former case
the source term was provided by tachyon fluctuations around a zero condensate
value.
In the sourceless case we rederived the same result by using CFT techniques
and demanding that the
relevant extra term in the $2d$ $\s$-model action (see \pertt) corresponds
to a marginal perturbation (which was argued to be exactly marginal).
In the case with sources the perturbation is not marginal by itself
but it produces
the necessary anomaly that cancels the term produced by the tachyon
fluctuations, so that the combined model stays conformal.
Moreover, the CFT method reveals that
these conditions have the same form to all orders in $\a'$.
We also gave explicit results in some important 4-dimensional cases
where the background geometry had an interpretation in terms of
exact CFT's. Further utilization of the CFT method is done
in appendix B (see (B.3)(B.4)).

{}From a string phenomenological point of view, the fact that random tachyon
fluctuations give rise to gravitational shock waves is an
important conclusion since the non-linear interactions of shock waves lead
to interesting formations \KPen,
including black holes (see, for instance, \DHooftII).
Questions of this nature should be further investigated.

It would also
be interesting to consider scattering of particles and strings in the
shock wave geometries we have obtained and in particular associate
the results (for instance the pole structure of the $S$-matrix \hoo)
with the CFT properties of the corresponding backgrounds.

\bs
\centerline{ \bf Acknowledgments }

I would like to thank A.A. Tseytlin for bringing to my attention ref. \DASA.

\vfill\eject

\appendix A {Useful tensors}

The non-vanishing \CS\ symbols corresponding to the metric \metrgll\
are
\eqn\crist{\eqalign{ & \G^u_{uu}= - {\Fpv\ov 2A}+ {\Apu\ov A} \ ,\qq
\G^u_{ij}= -{\gpv\ov 2A}\ h_{ij} \cr
& \G^v_{uu}= {\Fpu\ov 2A} + {F \Fpv \ov 2 A^2} - {F \Apu\ov A^2}\ ,\qq
\G^v_{uv} ={\Fpv \ov 2A} \ ,\qq \G^v_{ui} = {\Fpi\ov 2 A}\cr
& \G^v_{vv}= {\Apv\ov A} \ ,\qq \G^v_{ij}= \bl(-{\gpu \ov 2 A}
+{F \gpv\ov 2 A^2}\br)\ h_{ij} \cr
&\G^i_{uu}= -{1\ov 2 g}\ h^{ik} F_{,k}\ ,\qq \G^i_{uj}= {\gpu\ov 2 g}\ \d^i_j
\ , \qq \G^i_{vj}= {\gpv\ov 2 g}\ \d^i_j \cr
& \G^i_{jk}= \ha h^{il}( h_{lk,j} + h_{lj,k} - h_{jk,l}) \ .\cr }}
Using the above expressions we find that the non-vanishing components
of the Ricci tensor are (we substitute $F=-2 A f \d $ (see \metrgll))
\eqn\ricci{\eqalign{ &R_{uu}= {d-2\ov 2}\ \bl( {\gpu \Apu\ov g A} -
{g_{,uu} \ov g} + {\gpu^2 \ov 2 g^2} \br)
\ +\ {A\ov g}\ \d\ \triangle_{h_{ij}} f\ -\ {d-2\ov 2}\ {\gpv \ov g}\ \d' \ f
\cr
& \qq
 +\ \bl(2\ {A_{,uv}\ov A} - 2\ {\Apu \Apv\ov A^2} + {d-2\ov 2 g A}\ (\gpu \Apv
+ \gpv \Apu) \br)\ \d\ f  \cr
& \qq
 +\ 2\ \bl({A_{,vv}\ov A}- {\Apv^2\ov A^2} + {d-2\ov 2}\ { \gpv\Apv \ov g
A}\br)\
 \d^2\ f^2 \cr
&R_{uv}= \bl( {\Apu \Apv\ov A^2} - {A_{,uv}\ov A}
+ {d-2\ov 4}\ {\gpu\gpv\ov g^2} - {d-2\ov 2}\ {g_{,uv}\ov  g } \br) \cr
&\qq
+\ \bl({\Apv^2\ov A^2}- {A_{,vv}\ov A} -{d-2\ov 2}\ {\gpv\Apv\ov gA}\br)\
 \d\ f\cr
& R_{ui}= -\bl( {d-4\ov 2}\ {\gpv\ov g} + {\Apv\ov A} \br)\ \d\ \fpi  \cr
&R_{vv}= {d-2\ov 2}\ \bl({\gpv \Apv\ov gA} + {\gpv^2\ov 2 g^2}
- {g_{,vv}\ov g} \br) \cr
&R_{ij}= R^{(d-2)}_{ij}\ -\
\bl( {d-4\ov 2}\ {\gpu\gpv\ov gA}+ {g_{,uv}\ov A}\br)
\ h_{ij}\ -\ \bl({d-4\ov 2}\ {\gpv^2\ov gA} + {g_{,vv}\ov A}\br)\ h_{ij}\
 \d\ f \ .\cr }}
The above expressions for the \CS\ symbols and the Ricci tensor were
derived in \shock. For the reader's convenience we included them
in this paper as well.
The components of $D_{\m}D_{\n} \Phi$ with $\Phi=\Phi(u,v,x)$ are given by
\eqn\comp{\eqalign{
& D_u D_u \Phi = ( \Phi_{,uu}- {\Apu \ov A} \Phi_{,u}) -
{1\ov A}\ ( \Apu \Phi_{,v} + \Apv \Phi_{,u})\ \d \ f +  \Phi_{,v}\ \d'\ f\cr
& - \ {A\ov g}\ \d\ h^{ij} f_{,i} \Phi_{,j}
-   2 {\Apv \Phi_{,v} \ov A}\ \d^2 \ f^2 \cr
& D_u D_v \Phi = \Phi_{,uv} + {\Apv \Phi_{,v} \ov A}\ \d \ f \cr
&D_u D_i \Phi = \Phi_{,ui} -
{\gpu\ov 2g}\ \Phi_{,i}  + \Phi_{,v}\ \fpi\ \d \cr
& D_v D_v \Phi = \Phi_{,vv}  - {\Apv \Phi_{,v}\ov A} \cr
&D_v D_i \Phi = \Phi_{,vi} - {\gpv\ov 2g}\ \Phi_{,i} \cr
& D_i D_j \Phi = \Phi_{,ij} - \G^k_{ij} \Phi_{,k}  +
{1\ov 2 A}\ (\gpu \Phi_{,v} + \gpv \Phi_{,u})\ h_{ij}
 +  {\gpv \Phi_{,v} \ov A}\ h_{ij} \ \d\ f \ .\cr }}
We also compute
\eqn\hhmn{\eqalign{
& (H^2)_{uu}= {1\ov g^2} H_{uij} H_{ukl} h^{ki} h^{lj}
+ {4\ov g A} H_{uvi} H_{uvj} h^{ij} f \d \cr
& (H^2)_{uv}= {1\ov g^2} H_{uij} H_{vkl} h^{ki} h^{lj}
- {2\ov g A} H_{uvi} H_{uvj} h^{ij}  \cr
& (H^2)_{vv} = {1\ov g^2} H_{vij} H_{vkl} h^{ki} h^{lj} \cr
& (H^2)_{ui}= {1\ov g^2} H_{ujl} H_{imn} h^{jm} h^{ln}
-{2\ov g A}H_{uvj} H_{uim} h^{mj} - {4\ov A g} H_{uv j} H_{uim} h^{mj} f \d \cr
& (H^2)_{vi}= {1\ov g^2 } H_{vjl} H_{imn} h^{mj} h^{nl}
+{2\ov g A} H_{uvj} H_{vim} h^{mj} \cr
& (H^2)_{ij} = {1\ov g^2} H_{ikl} H_{jmn} h^{km} h^{ln}
-{2\ov A^2} H_{iuv} H_{juv} \cr
& \qq \ \ \ + {2\ov g A}
\bl(H_{imu} H_{jnv} + H_{imv} H_{jnu} + 2 H_{imv} H_{jnv} f\d \br) h^{mn}\ .
\cr } }



\appendix B { Shock waves on more general string backgrounds }

In this appendix we construct shock waves on more general than \metrd\
string backgrounds using the new method based on CFT techniques
that was introduced in section 2. Consider the string background
\eqn\megene{\eqalign{& ds^2 = 2\ A(u,v,x)\ dudv + E(u,v,x)\ du^2 +
g_{ij}(u,v,x)\ dx^i dx^j \cr
& B= 2 B_{ui}(u,v,x)\ du \wedge dx^i + B_{ij}(u,v,x)\ dx^i \wedge dx^j  \cr
&\Phi=\Phi(u,v,x)  \ .\cr}}
Let us add to the $\s$-model action corresponding to that a similar
to \pertt\ term
\eqn\ppee{ -2 \int d^2z\ \d(u) A(u,v,x) \l(u,v,x) \ \del u\bd u \ .}
Notice that because of the dependence of $\l(u,v,x)$ on the extra coordinates
$u, v$ this term cannot be obtained, in general,
via a simple shift of $v$ as \pertt.
Demanding that $\del u\bd u$ has dimension $(1,1)$ with respect to the
energy momentum tensor corresponding to \megene\ imposes the conditions
\eqn\ccaa{ \Apv = E_{,v} = g_{ij,v} = \Phi_{,v} = \l_{,v} = 0
\quad {\rm at}\ u=0\ .}
Notice that, among other differences with \conff, there is now a condition
on the function $\l(u,v,x)$ itself.
Demanding that $\l(u,v,x) \d(u)$ transforms like a function and adding the
source term (due, for instance, to stochastic tachyon fluctuations) we obtain
the linear differential equation
\eqn\condii{\eqalign{
& \triangle \l(0,v,x)  -  c(v,x) \l(0,v,x)  =  2\pi k\ \d^{(d-2)}(x-x') \cr
& c(v,x)\equiv {1\ov A} (\ha g^{ij} g_{ij,uv} +  \Phi_{,uv}) \ ,\cr }}
where all functions are computed at $u=0$ and the Laplacian is defined
as (cf.\laplaa)
\eqn\lapplaa{\triangle = {1\ov e^{\Phi} \sqrt{g} A}
\partial_i e^{\Phi}\sqrt{g} A \ g^{ij} \partial_j \ .}
It is easy to see that \ccaa\lapplaa\ reduce to \conff\condit\ when the
background \megene\ specializes to \metrd. In the cases where $E(u,v,x)$ is
zero at $u=0$, arguments similar to \anfa\uiuh\ show that the term \ppee\
corresponds to an exactly marginal perturbation.

An example of a background belonging to the more general class \megene\ is
given by
\eqn\wormm{\eqalign{& ds^2 = 2 V^{-1}(dudv + dzd\z)\ ,\quad V^{-1}= 2 C
+(uv+z\z)^{-1} \cr
& \Phi= \ln V\ ,\quad H_{\m\n\r} = 2 \e_{\m\n\r\l} \del^{\l}\Phi\ ,\cr}}
where $C$ is a constant. When $C=0$ this corresponds to the direct product CFT
of $SU(2)_k$ with a timelike boson having a background charge \ABEN\
or the Minkowski continuation of the `semi-wormhole' model,
with $N=4$ worldsheet supersymmetry, of \CHS.
It
is easy to show that \condii\ reduces to $\triangle \l=2\pi k\d^{(2)}(z-z')$,
i.e., the same as in the flat space case, with solution $\l=k \ln|z-z'|$.
This is not surprising since the Einstein metric
corresponding to \wormm\ is the \MI\ one.

It is important (in order to exclude any surprises)
to verify that \ccaa\condii\ also follow by requiring
that the $\b$-function equations are satisfied for the generic background
\megene.
Such a tedious computation has been explicitly performed and the result is
exactly \ccaa\condii.

\appendix C { Elements of stochastic calculus }

In this appendix we present some elementary facts of stochastic calculus that
are needed in order to prove \exptau-\souu. For more details
the reader should consult one of the many relevant books and review articles
in the literature (see, for instance, \gardiner).

A stochastic integral of a representative function $h(t)$
of a class of stochastic processes is defined as
\eqn\stodef{
I(t;t_0)= \int_{t_0}^t dw(t)\ h(t) \equiv \lim_{n\to \infty}
\sum_{k=1}^{n} h(\tau_k)\bl( w(t_k)-w(t_{k-1})\br) \ ,}
where $\tau_k$ is a point in the interval $[t_{k-1},t_k]$. The stochastic
variable $w(t)$ associated with a Brownian motion satisfies the properties
\eqn\brwt {\langle w(t) \rangle = w_0 \ ,\qq \langle w(t) w(s) \rangle =
(s-t_0 +w_0^2)\ \vartheta (t-s) + (t-t_0 +w_0^2)\ \vartheta (s-t) \ ,}
from which one easily proves that the stochastic differential $dw(t)$ obeys
\eqn\stodw{ \langle dw(t) \rangle =0 \ ,\qq \langle dw(t)dw(s) \rangle
 = \d(t-s) dt ds \ .}
The second relation states that $(dw(t))^2 = O(dt)$ and therefore $dw(t)$
should be treated as a differential of order $\ha$ in various algebraic
manipulations and Taylor expansions.
It turns out that the integral $I(t;t_0)$ is not independent of the choice
of the intermediate point $\tau_k \in [t_{k-1},t_k]$. The choice
$\tau_k=t_{k-1}$ corresponds to the {\it Ito calculus}.
For the average of two Ito integrals corresponding to two stochastic
functions $h_1(t)$, $h_2(t)$ the formula
\eqn\pooo{ \langle \int_{t_0}^{t} dw(t)\ h_1(t)
\int_{t_0}^{t} dw(s)\ h_2(s) \rangle =
\int_{t_0}^{t} dt\ \langle h_1(t)h_2(t) \rangle \ ,}
is extremely useful because it converts a double integration over the
stochastic variable $w(t)$ into a single ordinary integral. For the proof
of \pooo\ the definition \stodef\ and \brwt\ should be used together with the
crucial assumption that the stochastic functions $h_1(t)$, $h_2(t)$
are independent of $w(s)$ for $t< s$, namely, that
\eqn\iidd{ \langle h_i(t) w(s) \rangle = 0, \ \ i=1,2\ ,
\quad {\rm if}\ t<s \ .}
Obviously, if $h_1(t)$, $h_2(t)$
are deterministic functions there is no need to
take the average on the right hand side of \pooo. This is the case
in the derivation of \souu\ in section 2.

\listrefs
\end